\documentclass[twocolumn,showpacs,preprintnumbers,amsmath,amssymb]{revtex4}

\usepackage{amsmath}
\usepackage{graphicx}
\usepackage{calc}
\usepackage{bm}

\begin{document}
\title{Unusual anisotropy of inplane field magnetoresistance in ultra-high mobility SiGe/Si/SiGe quantum wells}
\author{M.~Yu. Melnikov}
\affiliation{Institute of Solid State Physics, Chernogolovka, Moscow District 142432, Russia}
\author{V.~T. Dolgopolov}
\affiliation{Institute of Solid State Physics, Chernogolovka, Moscow District 142432, Russia}
\author{A.~A. Shashkin}
\affiliation{Institute of Solid State Physics, Chernogolovka, Moscow District 142432, Russia}
\author{S.-H. Huang}
\affiliation{Department of Electrical Engineering and Graduate Institute of Electronics Engineering, National Taiwan University, Taipei 106, Taiwan, and\\ National Nano Device Laboratories, Hsinchu 300, Taiwan}
\author{C.~W. Liu}
\affiliation{Department of Electrical Engineering and Graduate Institute of Electronics Engineering, National Taiwan University, Taipei 106, Taiwan, and\\ National Nano Device Laboratories, Hsinchu 300, Taiwan}
\author{S.~V. Kravchenko}
\affiliation{Physics Department, Northeastern University, Boston, Massachusetts 02115, USA}

\begin{abstract}
We find an unusual anisotropy of the inplane field magnetoresistance with higher resistance in the parallel orientation of the field, $B_\parallel$, and current, $I$, in ultra-high mobility SiGe/Si/SiGe quantum wells. The anisotropy depends on the orientation between the inplane field and current relative to the crystallographic axes of the sample and is a consequence of the ridges on the quantum well surface. For the parallel or perpendicular orientations between current and ridges, a method of converting the magnetoresistance measured at $I\perp B_\parallel$ into the one measured at $I\parallel B_\parallel$ is suggested and is shown to yield results that agree with the experiment.
\end{abstract}
\maketitle

\section{Introduction}

Magnetoresistance of two-dimensional (2D) electron systems in inplane magnetic fields is normally caused by spin and orbital mechanisms. For a thin 2D electron system with thickness much smaller than the magnetic length, the inplane magnetic field only affects the degree of the spin polarization. The resistance, $R$, increases with $B_\parallel$ until the full spin polarization is reached and saturates after that \cite{simonian1997,pud,ok,vit,sh,DePoortere2002,lai2005,gao2006}, which is explained in terms of the dependence of the screening of the random potential on the degree of spin polarization \cite{dolg}. Being a spin effect, the parallel field-induced magnetoresistance should not depend on the mutual orientation of the field $B_\parallel$ and current $I$ \cite{pud1}. In the opposite case of a thick 2D electron system, the orbital effect comes into play and may even become dominant. In the presence of the magnetic field $B_\parallel=(0,B_\text{y},0)$, the electron wavefunction in $z$-direction perpendicular to the interface depends on the $k_\text{x}$ component of the Fermi wavevector, which leads to an increase of the effective mass in $x$-direction and deformation of the Fermi surface \cite{JPCM}. The resistance increases monotonically with magnetic field and becomes anisotropic, where the magnetoresistance for $I\perp B_\parallel$ is higher than for $I\parallel B_\parallel$. The effect has been observed in the 2D electron system in GaAs/AlGaAs heterostructures \cite{shay,Zhu}. Note that in an anomalously wide quantum well, parallel magnetic field may even cause changes in the topology of the Fermi surface, which are manifested by additional features on the magnetoresistance \cite{shay1}.

Another mechanism of the inplane field magnetoresistance is realized in 2D electron systems that are distinguished by a corrugated surface \cite{bykov,SST}. In this case, the mainly parallel magnetic field creates a nonuniform component of the field in the direction perpendicular to the surface, leading to a positive magnetoresistance.

Recently, unique SiGe/Si/SiGe quantum wells have been created \cite{lu1,lu2,lu3,mel}. The electron spectrum in these samples is the same as that in (100) silicon metal-oxide-semiconductor field-effect transistors (MOSFETs), but the electron mobility is two orders of magnitude higher. These 2D systems are a promising candidate for studying the energy spectrum and transport properties of strongly interacting electrons \cite{arxiv}. Anisotropy of the magnetoresistance opposite to that expected for the finite thickness of the quantum well was reported for SiGe/Si/SiGe quantum wells \cite{oka}. However, the cause of this anisotropy remained unclear.

\begin{figure}[b]
\scalebox{0.34}{\includegraphics[angle=0]{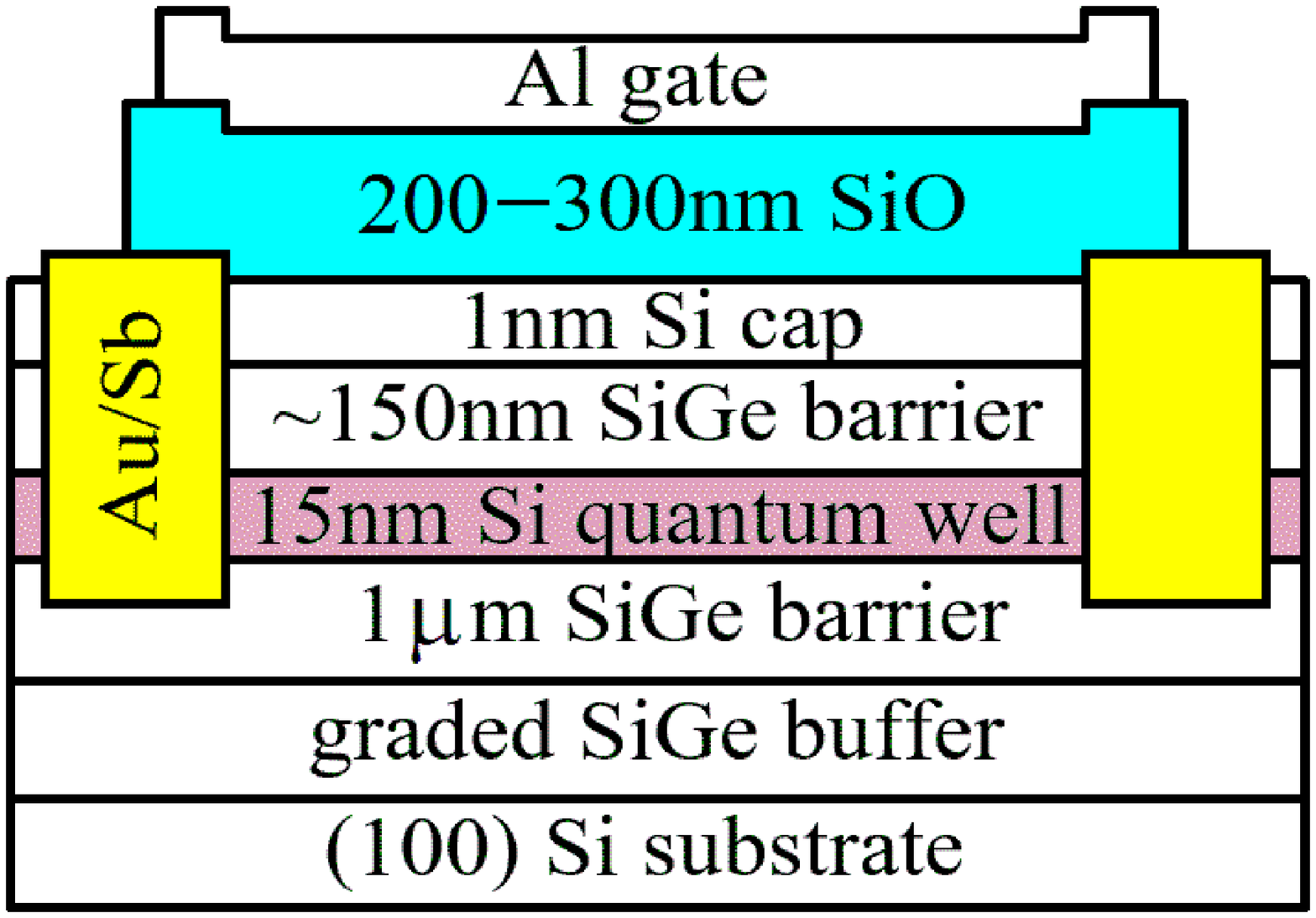}}
\caption{\label{fig1} Illustration of the sample structure.}
\end{figure}

In this paper, we report studies of the magnetoresistance of ultra-high mobility SiGe/Si/SiGe quantum wells in inplane magnetic fields up to and exceeding the fields for complete spin polarization of the electrons. We find an unusual resistance anisotropy with higher $R(B_\parallel)$ in the parallel orientation of the field and current that is determined by orientation of the current relative to the crystallographic axes of the sample. Using atomic-force microscope studies, we show that the observed anisotropy is related to the morphology of the interface, namely, the ridges on the quantum well surface. For the parallel or perpendicular orientations between current and ridges, a method of converting the magnetoresistance measured at $I\perp B_\parallel$ into the one measured at $I\parallel B_\parallel$ is suggested. This is in good agreement with the experiment.

\begin{figure}
\scalebox{0.4}{\includegraphics[angle=0]{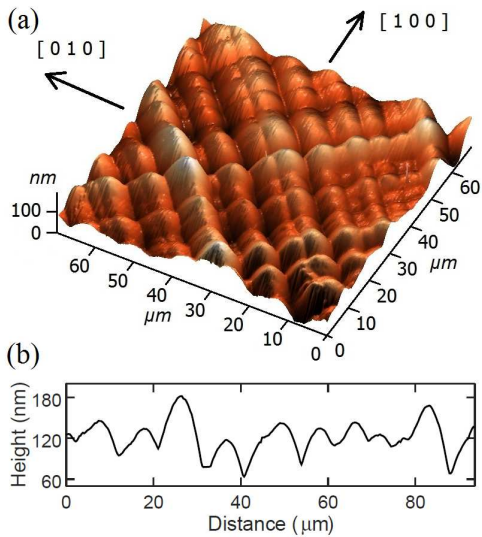}}
\caption{\label{fig2} (a) Atomic-force microscope image of the wafer surface with a resolution of $256\times 256$ dots. (b) The line profile in the [-110] direction across the right corner of the image.}
\end{figure}

\section{Samples and Methods}

We used SiGe/Si/SiGe quantum wells grown in an ultra-high vacuum chemical-vapor deposition chamber on (001) Si substrates \cite{lu1,lu2,lu3}. The cross-section of the sample is shown schematically in Fig.~\ref{fig1}. The approximately 15~nm wide silicon quantum well is sandwiched between Si$_{0.8}$Ge$_{0.2}$ potential barriers. The samples were patterned in Hall-bar shapes with distance between the potential probes of 150~$\mu$m and width of 50~$\mu$m using standard photo-lithography (for more details, see Ref.~\cite{melnikov2014}). Electric contacts to the 2D layer consisted of AuSb alloy, deposited in a thermal evaporator in vacuum and annealed for 3 to 5 minutes at 440$^\text{o}$C in N$_2$ environment. An approximately 200 -- 300~nm thick layer of SiO was then deposited in a thermal evaporator and an aluminum gate was deposited on top of the SiO. No additional doping and no mesa etching were used; the 2D electron gas was created by applying positive voltage to the gate in a way similar to silicon MOSFETs. Four samples made from two different pieces (SiGe2 and SiGe3) of the same wafer were studied. In samples SiGe2-I and SiGe2-II, the long side of the Hall bar corresponded to the direction of current parallel to the [-110] and [110] crystallographic axes, respectively. The results obtained on these two samples were similar. In samples SiGe3-I and SiGe3-II, the long side of the Hall bar was parallel to the [100] axis. The maximum mobility was about $240$~m$^2$/Vs at electron density $n_\text{s}\approx10^{11}$~cm$^{-2}$ \cite{mel}.

Measurements were carried out in an Oxford TLM-400 dilution refrigerator at a temperature $T\approx0.03$~K. Magnetoresistance was measured with a standard four-terminal lock-in technique in a frequency range 1--11~Hz in the linear regime of response (the measuring current was varied in the range between 0.5 and 4~nA). To improve the quality of contacts and increase the electron mobility, a saturating infra-red illumination of the samples was used; after the resistance decreased and saturated with time, the illumination was turned off. This did not affect the electron density at a fixed gate voltage.

\section{Experimental Results}

In Fig.~\ref{fig2}, an atomic-force microscope image of the wafer surface is shown. The tapping mode was used, and the results of scanning in [100] and [010] directions were checked to be consistent with each other. For our case the main artifacts distorting the image were the line defects in scanning direction, the portion of line scans with defects being small. Each of them was replaced by the superposition of the nearest lines obtained by means of a cube approximation using a standard program for treatment of atomic-force microscope images, which resulted in Fig.~\ref{fig2}. Ridges on the wafer surface are coincident with the [110] and [-110] crystallographic axes as determined from X-ray diffraction analysis. In both directions, the characteristic period of the ridges is $\sim10$~$\mu$m and the depth is $\sim50$~nm. Note that such a regular cross-hatch pattern is known for the SiGe/Si heterostructures \cite{SST97}. By etching the wafer to just below the Si quantum well, we have verified that ridges on the surface of SiGe barrier are virtually the same as the ridges on the wafer surface.

\begin{figure}
\scalebox{0.34}{\includegraphics[angle=0]{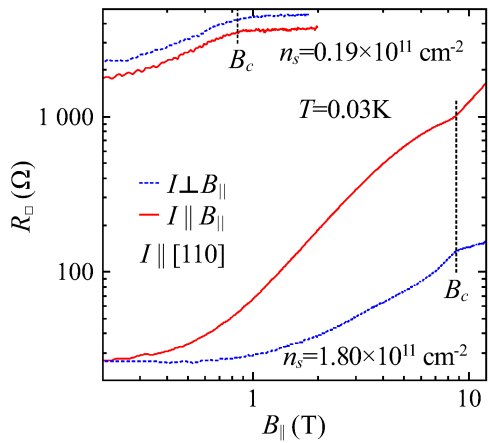}}
\caption{\label{fig3} Parallel-field magnetoresistance of sample SiGe2-II. Solid lines show the data obtained for $I\parallel B_\parallel$ and dotted lines correspond to $I\perp B_\parallel$. The upper solid line is vertically shifted for clarity by dividing by the factor of 1.2.}
\end{figure}

Parallel-field magnetoresistance of sample SiGe2-II for $I\parallel [110]$ at different orientations between field and current is shown in Fig.~\ref{fig3}. At low electron densities, the magnetoresistance curves are independent of whether the current is parallel or perpendicular to the magnetic field. The resistance increases with magnetic field by approximately a factor of 2 and saturates at a magnetic field of the complete spin polarization, $B_\text{c}$. The ratio $R(B_\text{c})/R(0)$ is known to be dependent on the scattering potential \cite{SSC}. An increase in the electron density leads to appreciable changes in the magnetoresistance measured in the $I\parallel B_\parallel$ configuration. The dependences no longer saturate at the field of the complete spin polarization, and the relative magnitude of the magnetoresistance becomes much larger. In contrast, the magnetoresistance for $I\perp B_\parallel$ behaves qualitatively the same way as at low $n_\text{s}$, even though the ratio $R(B_\text{c})/R(0)$ grows significantly. Similar results are obtained in sample SiGe2-I in which the direction of current is parallel to [-110] axis.

\begin{figure}
\scalebox{0.34}{\includegraphics[angle=0]{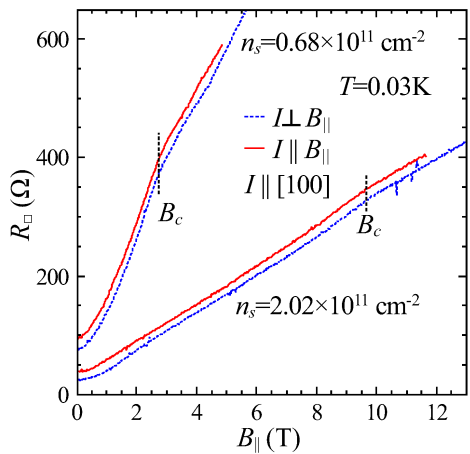}}
\caption{\label{fig4} Parallel-field magnetoresistance of sample SiGe3-I. Solid lines show the data obtained for $I\parallel B_\parallel$ and dotted lines correspond to $I\perp B_\parallel$. The solid lines are shifted upward by 15~$\Omega$ for clarity. To determine the field $B_c$, the monotonic change in the resistance with $B_\parallel$ was subtracted.}
\end{figure}

In Fig.~\ref{fig4}, we show the magnetoresistance data for sample SiGe3-I in which the current is directed at 45$^\text{o}$ relative to the ridges, along [100] axis. In contrast to the data from Fig.~\ref{fig3}, the magnetoresistance in the whole range of electron densities between 0.68 and $2\times10^{11}$~cm$^{-2}$ is the same for both mutual orientations of the field and current. The absence of the anisotropy indicates that the orbital effects due to the finite thickness of the 2D electron system are small in our case. The identical behavior of the magnetoresistance is observed in sample SiGe3-II.

\begin{figure}
\scalebox{0.6}{\includegraphics[angle=0]{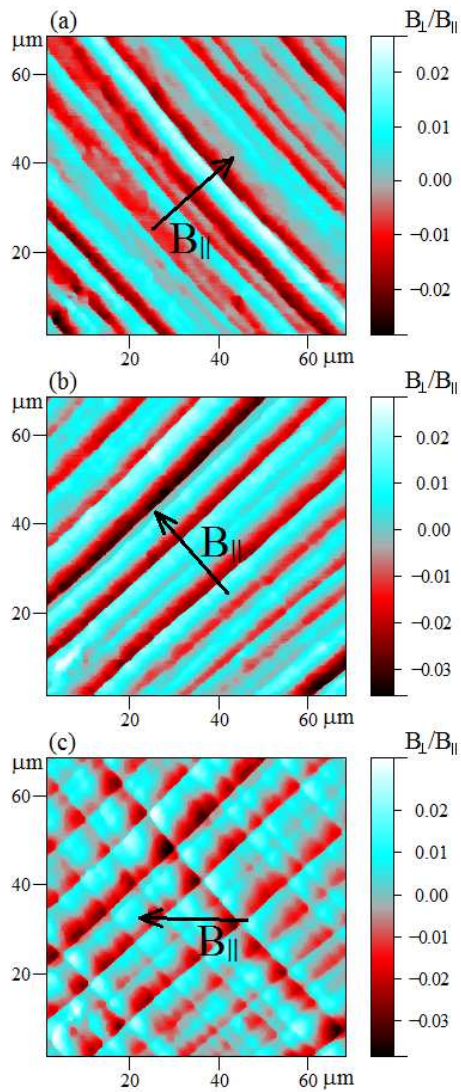}}
\caption{\label{fig5} Relative value of the perpendicular component of the magnetic field for the part of the sample displayed in Fig.~\ref{fig2} for three different orientations of the magnetic field, shown by arrows.}
\end{figure}

Thus, the shape and anisotropy of the magnetoresistance in our samples depend crucially on the mutual orientation between the current and the direction of the ridges. This points to ridges being the main cause for the anisotropy.

\section{Discussion}

The likely reason for the influence of ridges on the magnetoresistance can be the existence of a local out-of-plane field component, $B_\perp$, of the mainly inplane magnetic field that is directly related to the structural anisotropy. The component $B_\perp$ arises because the angle between the corrugated surface and the mainly inplane magnetic field is not everywhere zero. For an electron gas on $Z(x,y)$ surface this component is equal to

\begin{equation}
B_\perp(x,y)=\frac{-B_\text{x}\frac{\partial Z}{\partial x}-B_\text{y}\frac{\partial Z}{\partial y}}{\sqrt{\left(\frac{\partial{Z}}{\partial{x}}\right)^2+\left(\frac{\partial{Z}}{\partial{y}}\right)^2+1}},\label{1}
\end{equation}
where the values of $\partial Z/\partial x$ and $\partial Z/\partial y$ can be obtained directly from the results of Fig.~\ref{fig2}. The calculated out-of-plane component $B_\perp$ of the magnetic field in the part of the sample displayed in Fig.~\ref{fig2} is shown in Fig.~\ref{fig5} for three directions of the inplane magnetic field. As seen from Figs.~\ref{fig5}(a) and (b), the distribution of the components $B_\perp$ forms similar stripes perpendicular to the direction of $B_\parallel$, which explains the similar results obtained on samples SiGe2-I and SiGe2-II.

\begin{figure}
\scalebox{0.44}{\includegraphics[angle=0]{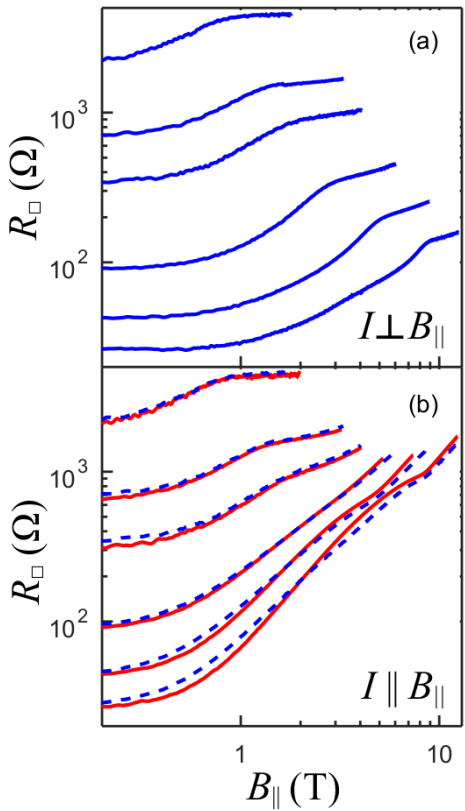}}
\caption{\label{fig6} Magnetoresistance of SiGe2-II for $I\parallel[110]$ in both orientations of the field and current at $T\approx0.03$~K and $n_\text{s}=0.19, 0.30, 0.40, 0.64, 1.10, 1.80\times10^{11}$~cm$^{-2}$ (from top to bottom; solid lines). Also shown by dashed lines in (b) is the magnetoresistance converted from the data in (a) using Eq.~(\ref{2}).}
\end{figure}

Knowing the distribution of the perpendicular component of the magnetic field (Figs.~\ref{fig5}(a) and (b)), it is easy to estimate the magnetoresistances for different orientations of the current and field. Indeed, if $I\perp B_\parallel$, the stripes are parallel to the direction of the current. In the neighboring stripes, the directions of the electrons' drift due to the Hall effect are opposite, and electric charge accumulates between the stripes. In other words, the stripes can be considered as narrow Hall bars in zero-on-average $B_\perp$, and the resulting magnetoresistance is only caused by spin polarization for a thin 2D electron system. However, if $I\parallel B_\parallel$, the stripes are perpendicular to the current, and the Hall currents in the neighboring stripes are balanced. Therefore, the sample geometry is equivalent to the Corbino one. Using the formula for the dissipative conductivity in a perpendicular magnetic field, $B$: $\sigma_{xx}=\sigma(B=0)/(1+\mu^2B^2)$ (where $\mu$ is the mobility), we get for our case

\begin{equation}
R(I\parallel B_\parallel)=R(I\perp B_\parallel)\left(1+\frac{\mu^2B_\parallel^2}{L}\int\limits_0^L\frac{B_\perp^2}{B_\parallel^2}dr\right),\label{2}
\end{equation}
where the integration is done in the direction perpendicular to the stripes, $L$ is the distance between potential probes, and the electron mobility at each $B_\parallel$ is presumed to have some averaged value $\mu=1/n_\text{s}eR_\square(I\perp B_\parallel)$. Equation~(\ref{2}) allows one to calculate the magnetoresistance in the configuration $I\parallel B_\parallel$ using the known magnetoresistance for $I\perp B_\parallel$. We have verified that the images obtained for different positions of the scanning window of the atomic-force microscope on the sample are similar. Assuming that the distribution of the perpendicular component of the magnetic field is the same over the sample, we use the value of the integral in Eq.~(\ref{2}), averaged over an area of about 1,000 $\mu$m$^2$, for comparison with the experiment.

Magnetoresistances measured at the same electron densities for both orientations of the field and current are shown by solid lines in Fig.~\ref{fig6}. The dashed lines in Fig.~\ref{fig6}(b) correspond to the magnetoresistance converted from the data in Fig.~\ref{fig6}(a) using Eq.~(\ref{2}). As seen from Fig.~\ref{fig6}(b), the converted magnetoresistance is in good agreement with the experiment. According to Eq.~(\ref{2}), the degree of the anisotropy is determined by electron mobility. The anisotropy should disappear at low electron densities due to a drop in the mobility, which is consistent with the data in Fig.~\ref{fig3}. At high electron densities, when the second term in brackets in Eq.~(\ref{2}) is dominant, one obtains $R(I\parallel B_\parallel)\propto 1/R(I\perp B_\parallel)$ so that the shoulder on the magnetoresistance for $I\parallel B_\parallel$ should get inverted (see traces at the two highest electron densities in Fig.~\ref{fig6}(b)).

In the configuration of Fig.~\ref{fig5}(c), the magnetoresistances in both perpendicular and parallel orientations of the field $B_\parallel$ and current $I$ should be identical ({\it cf.} Fig.~\ref{fig4}), because the distribution of the perpendicular component of the magnetic field remains the same on average after the 90$^\text{o}$ change in the direction of $B_\parallel$.

\section{Conclusions}

In conclusion, we have found that the presence of the unusual anisotropy of the parallel-field magnetoresistance in ultra-high mobility SiGe/Si/SiGe quantum wells with respect to mutual orientation between the field and current is determined by orientation of the current relative to the crystallographic axes of the sample and is a consequence of the ridges on the quantum well surface. A quantitative model is presented that maps the magnetoresistance for field perpendicular to the current to that for field parallel to current.

\section{Acknowledgments}

We are grateful to Don Heiman for helpful suggestions. The ISSP group was supported by RFBR grants \# 15-02-03537 and 16-02-00404, Russian Academy of Sciences, and Russian Ministry of Science. The NTU group was supported by MOST, Taiwan, under contract Nos. 105-2622-8-002-001, 103-2221-E-002-253-MY3, and 103-2221-E-002-232-MY3. SVK was supported by NSF Grant No.\ 1309337 and BSF Grant No.\ 2012210.

\end{document}